\documentclass{article}
\usepackage{nips2003e,times}
\usepackage{graphicx}
\usepackage{subfigure}

\title{Centralized reward system gives rise to fast and efficient work sharing for intelligent Internet agents lacking direct communication}

\author{
Zsolt Palotai \\
Department of Information Systems\\
E\"otv\"os University, Budapest, Hungary\\
\texttt{zspalotai@vnet.hu} \\
\And
S\'{a}ndor Mandusitz \\
Department of Information Systems\\
E\"otv\"os University, Budapest, Hungary\\\
\texttt{santi@inf.elte.hu} \\
\AND
Andr\'as L\H{o}rincz\thanks{ corresponding author} \\
Department of Information Systems\\
E\"otv\"os University, Budapest, Hungary\\
\texttt{lorincz@inf.elte.hu} \\
}

\begin{document}

\maketitle

\begin{abstract}
WWW has a scale-free structure where novel information is often
difficult to locate. Moreover, Intelligent agents easily get
trapped in this structure. Here a novel method is put forth, which
turns these traps into information repositories, supplies: We
populated an Internet environment with intelligent 'news
foragers'. Foraging has its associated cost whereas foragers are
rewarded if they detect not yet discovered novel information. The
intelligent `news foragers' crawl by using the estimated long-term
cumulated reward, and also have a finite sized memory: the list of
most promising supplies. Foragers form an artificial life
community: the most successful ones are allowed to multiply, while
unsuccessful ones die out. The specific property of this community
is that there is no direct communication amongst foragers but the
centralized rewarding system. Still, fast division of work is
achieved.

\end{abstract}

\section{Introduction}
\label{s:intro}

The largest source of information today is the World Wide Web. The
number of documents may reach 10 billion soon; the number of
documents changing on a daily basis is also enormous. The
ever-increasing growth presents a considerable challenge in
finding \textit{novel information} on the web.

WWW has a scale-free structure
\cite{barabasi00scalefree,Kleinberg01structure}: a graph is a
scale-free network if the number of incoming (or outgoing or both)
edges follows a power-law distribution ($P(k) \propto
k^{-\gamma}$, where $k$ is integer, $P(k)$ denotes the probability
that a vertex has $k$ incoming (or outgoing or both) edges and
$\gamma >0$). The direct consequence of the scale-free property is
that there are numerous URLs or sets of interlinked URLs, which
have a large number of incoming links. Intelligent web crawlers
can be easily trapped at the neighborhood of such junctions
\cite{diligenti00focused,kokai02learning,Lorincz02intelligent,rennie99using}.

We have developed a novel artificial life (Alife) method with
intelligent individuals (agents) to detect `breaking news' type
information on a prominent and vast WWW domain. We turned to Alife
to achieve efficient division of labor under minimized
communication load between individuals. For reviews on relevant
evolutionary theories, see, e.g.,
\cite{Clarck00dynamic,Fryxell98individual}. See
\cite{Csanyi89evolutionary,kampis91selfmodifying} on the dynamics
of self-modifying systems. See, e.g., \cite{Kennedy01swarm} for a
review on emerging intelligence in `swarm' communities. Our agents
crawl by estimating the long-term cumulated reward using
reinforcement learning. (For a review on reinforcement learning,
see, e.g., \cite{Sutton98Reinforcement}.) The estimation uses
function approximation and temporal difference learning. Inputs to
the function approximator are provided by probabilistic
term-frequency inverse document-frequency (PrTFIDF) classifier
\cite{joachims97probabilistic} using text classes formed by a
well-known text clustering method \cite{Boley98principal}.

First, the applied methods are described in details (Section
\ref{s:methods}). Experimental results are provided in Section
\ref{s:results} followed by a discussion and a summary (Section
\ref{s:disc}).

\section{Methods}
\label{s:methods}

In our model, individuals are \textit{foragers}, `populating' a
continuously changing world, where the rate of the emergence of
new resources is limited.

\begin{figure}[t!]
  \centering
   \subfigure[Power law]
      {
         \includegraphics[width=6cm]{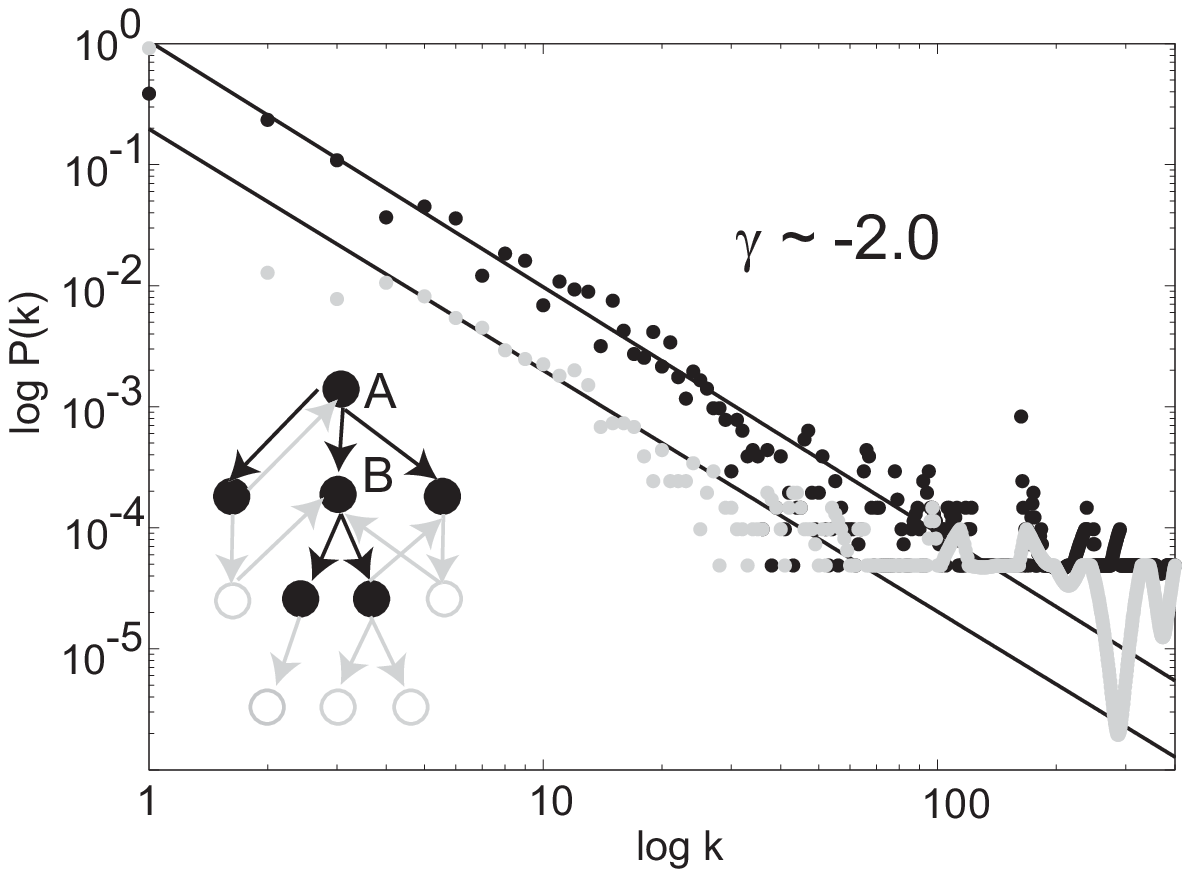}\label{f:fig1a}
      }
   \subfigure[Reinforcement]
      {
         \includegraphics[width=6cm]{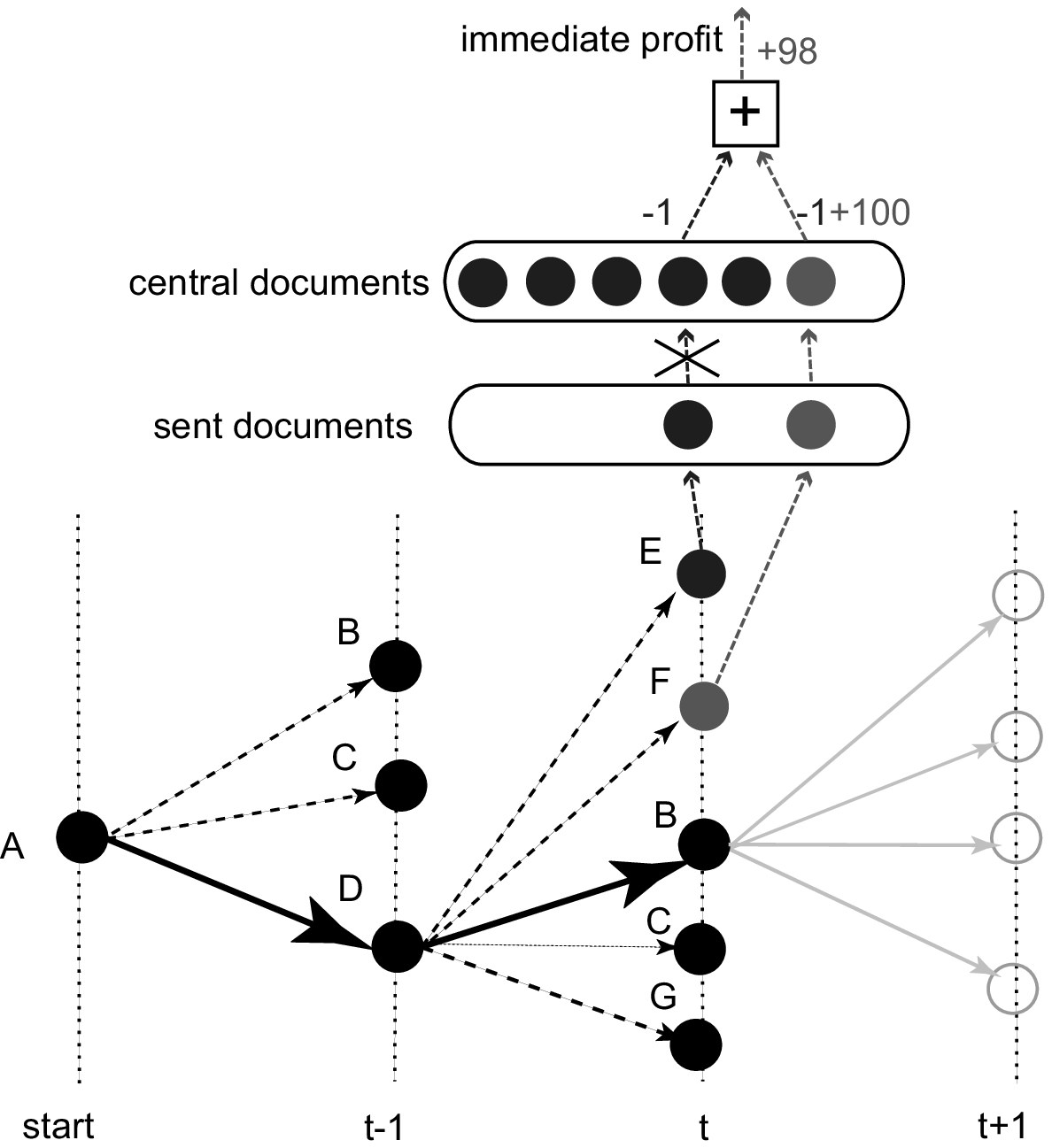}\label{f:fig1b}
      }
  \caption{\textbf{Scale-free properties of the Internet domain
  and reward system}
  \newline\textbf{(a):} Log-log scale distribution of
the number of (incoming and outgoing) links of all URLs found
during the time course of investigation.  Horizontal axis: number
of edges ($\log k$). Vertical axis: relative frequency of number
of edges at different URLs ($\log P(k)$). Black (gray) dots:
incoming (outgoing) edges of URLs. Slope of the straight lines
$-2.0 \pm 0.3$. \textit{Inset}: method of downloading. Black
(gray) link: (not) in database. Solid (empty) circle: (not) in
database. \textbf{(b):} Central reinforcement system. Empty
(solid) circles: (not) novel documents. Positive (negative)
numbers: reward and profit (cost). Vertical dashed lines:
consecutive time steps. Dots on the $(t+k)^{th}$ dashed line:
documents available at time step $t+k-1$.}\label{f:fig1}
\end{figure}

\textbf{Environment.} The domain of our experiments (one of the
largest news sites), as most WWW domains, was scale-free according
to our measurements. The distributions of both incoming and
outgoing links show a power distribution (Fig. \ref{f:fig1a}). The
inset shows the links and documents investigated by a forager when
it takes a step. The distributions, shown in the figure,
correspond to links investigated by the foragers. The news forager
visits URL `A', downloads the not-yet visited part of the
environment (documents of URLs, which URLs have not been visited
yet and are linked from URL `A'). Downloading is followed by a
decision, URL `B' is visited, downloading starts, and so on. The
series `A', `B', `C', ...  is called \textit{path}.

\textbf{The foragers.} Foragers have two components of long-term
memory. The first component serves decision making and determines
the \emph{behavior} of the forager. Foragers with different
behaviors may make different decisions at the same URL. The second
component is a finite-sized memory, which contains procedures.
This `procedural memory' codes how to escape from a trap. Here, a
procedure is simply a switching to a new URL, a new
\textit{starting point}. The forager searches for a given number
of steps. Steps are taken according to the forager's behavior.
After the given number of steps are taken, a procedure, which is
selected randomly from the procedural memory, solves the problem
of being trapped.

At each visited page, the forager downloads the neighboring
documents and determines whether a document has a time stamp of
the current (actual) date. If it does, this document is sent to
the center (these documents will be referred as `sent' documents).

\textbf{Reward system.} Foragers are searching for `food', which
is novel news. A central reinforcing unit administers positive and
negative rewards. Positive reward is delivered only to the first
sender of a given news item only if the document's time stamp is
not older than a day according to GMT. Then reward $c_+$ ($\,=100$
in arbitrary units (a.u.)) is `provided'. Each sending of a
document, costs $c_-$ ($=-1$ a.u.) for the forager. The
(immediate) profit is the difference of rewards and costs at any
given step. Figure \ref{f:fig1b} shows the reinforcement of a step
in an example path: at start (time $t-2$), the agent is at URL
`A', where documents of neighboring  URLs `B', `C' and `D' are
downloaded. URL `D' is visited next. Documents of URLs `E', `F'
and `G' are downloaded. Document of URL `G' has an obsolete date.
Documents of URLs `E' and `F' are sent to the center.  Document of
URL `F' is novel to the center, so it is rewarded. In turn, profit
98 is received by the forager.The forager maintains a list of
neighbors of visited URLs, called \textit{frontier}. It can only
visit URLs of the \textit{frontier}.

\textbf{Long term cumulated profit. (LTP)} Immediate profit is a
myopic characterization of an URL. Foragers \textit{behave}
intelligently by following the \textit{policy} that maximizes the
expected LTP instead of the immediate profit. Policy and profit
estimation are interlinked concepts: profit estimation determines
the policy, whereas policy influences choices and, in turn, the
expected LTP. (For a review, see \cite{Sutton98Reinforcement}.)
Here, choices are based on the greedy LTP policy: The forager
visits the URL, which belongs to the \textit{frontier} and has the
highest estimated LTP. Visited URLs form a path and each path is
limited to 100 steps.

Each forager has a $k(=50)$ dimensional PrTFIDF text classifier
\cite{joachims97probabilistic} generated on a previously
downloaded portion of the Geocities database. 50 clusters was
created by Boley's clustering algorithm \cite{Boley98principal}
from the downloaded documents. The PrTFIDF classifier was trained
on these clusters plus an additional one representing general
texts from the internet. The PrTFIDF outputs were non-linearly
mapped to interval [-1,+1] by a hyperbolic-tangent function. The
classifier was applied to reduce the texts to a small dimensional
representation. When the forager visits URL `A' displaying
document $d_a$, the output vector of the classifier is
$\mathbf{s_a}=(s_a(1),\ldots , s_a(k))$. (The $(k+1)^{th}$ output
was dismissed.)

A linear approximator is used for LTP estimation. It encompasses
$k$ parameters, the \textit{weight vector}
$\mathbf{w}=(w(1),\ldots , w(k))$. The LTP of document $d_a$ is
estimated as the scalar product of $\mathbf{s_a}$ and
$\mathbf{w}$: $L(d_a)=\sum_{i=1}^k w(i)s_a(i)$. This weight vector
is the first component of a forager's long-term memory. The weight
vectors are tuned by temporal difference learning
\cite{sutton88learning}: let us denote the document to be visited
next by $d_n$, the output of the classifier by $\mathbf{s_n}$ and
the estimated LTP of the document by $L(d_n) = \sum_{i=1}^k
w(i)s_n(i)$. Assume that leaving the actual document $d_a$ and
arriving to the next document, we have immediate profit $r_n$. Our
estimation is perfect if $L(d_a)=L(d_n)+r_n$. Future profits are
typically discounted in such estimations: $L(d_a)=\gamma
L(d_n)+r_n$, where $0 < \gamma < 1$. In turn, the error of value
estimation is

$$\delta(a,n) = r_n + \gamma L(d_n) - L(d_a).$$

\noindent Throughout the simulations $\gamma =0.9$ was used. At each actual step $d_a
\rightarrow d_n$ the weights of the value function were tuned to
decrease the error of value estimation for the visited documents.
This estimation error was used to correct the parameters: the
$i^{th}$ component of the weight vector $w_i$ was corrected by

 $$\Delta w_i = \alpha \,\delta(a,n) \, s_a(i)$$

\noindent with $\alpha=0.1$ and $i=1, \ldots , k$.

\textbf{URL lists and decisions.} The second component of
foragers' long-term memory, the procedural memory, is the
\textit{weblog}, which is a list of URLs with items less than or
equal to 100. \textit{Starting points} are the first 10 elements
of the weblog. At the start of a path the forager makes a random
choice amongst starting points and visits that URL. After a path
is finished the forager selects a new starting point for the next
path.

For an URL `A', the cumulated profit is the sum of immediate
profits collected during the path after visiting URL `A'. Denoting
the cumulated profit of URL `A' by $R_{path}(A)$, when a path is
completed, the \textit{value of the URL `A'}, denoted by $V(A)$ is
estimated as follows: $$V_{new}(A)=(1-\beta)V_{old}(A) + \beta
R_{path}(A)$$ where $\beta$ was set to $0.3$. If URL `A' did not
have a value before, then $V_{new}(A)$ is set to $R_{path}(A)$.
These values are then used to update the weblog after each path.
URLs are ordered by decreasing value and the list is clipped at
the $100^{th}$ URL to form the new weblog.

The forager also maintains two short-term memory lists during each
path. One of the lists contains URLs visited during the path to
avoid loops. The other list is the \textit{frontier}, which
contains the URL's of pages directly accessible from the visited
pages, excluding the visited URLs themselves. Forager selects its
next step from this list. If the list is empty or the actual path
has length of 100 steps then this path is finished.

\textbf{Multiplying by bipartition and extinction.} Every forager
has a value of 100 at start. The value is reduced by 0.05 for each
document sent to the center and increased by 1 if a sent document
is reinforced. Once the forager's value reaches 200, the forager
multiplies by bipartition and value 100 is assigned to both
descendants. The weblog of the parent is randomly separated into
two 50 element lists. The original weight vector of the parent and
the partial weblogs are passed on to the descendants. On the other
hand, if the forager's value hits 0 then it dies out.

\textbf{Foraging periods.} Foragers run sequentially in a
prescribed order for approximately equal time intervals on one PC (the analysis of the last visited URL (donwloading neighbouring URLs and LTP estimation) is allowed). The foraging period is the time interval while all foragers run once. Unfinished paths are continued in the next run.

\section{Experimental results}
\label{s:results}

Multiple two-week long experiments were conducted between
September and November 2002. Apart from short breaks, monitoring
was continuous for each two week period. Most of the figures in
this article represent the November 5-21, 2002 time period. The
parameters of this experiment are representative to the entire
series: The number of downloaded, sent and reinforced documents
was 1,090,074, 94,226 and 7,423, respectively. The used Internet
bandwidth (1Mbps on average) was almost constant, decreasing
slowly by 10 \% towards the end of the experiment. The number of
foragers increased from 2 to 22. Experiments were run on a single
computer. Within each foraging-period, the allocated time of every
forager was 180 s. Some uncertainty ($50\pm 30$s) arose, because
foragers were always allowed to complete the analysis of the last
URL after their allocated time expired. The net duration of a 100
step path was ~350s.

\subsection{Time lag and multiplication} \label{s:efficiency}

\begin{figure}[t!]
 \centering
  \subfigure[Forager efficiency]
      {
         \includegraphics[width=8cm]{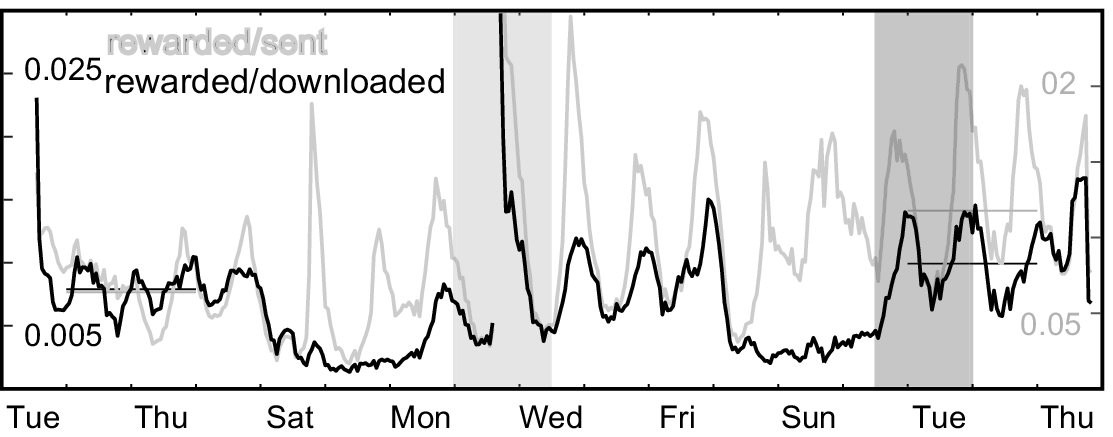}\label{f:fig2a}
      }
   \subfigure[Population value vs. time]
      {
         \includegraphics[width=88mm]{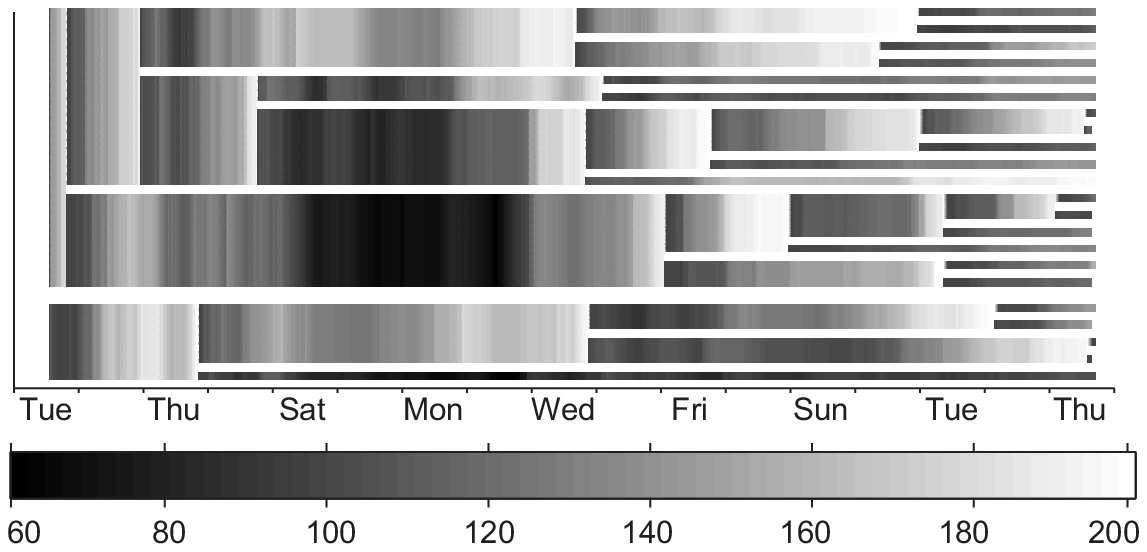}\label{f:fig2b}
      }
\caption{\textbf{Experimental results}\newline\textbf{(a):} The
rate of sent and rewarded documents showed daily oscillations.
Lighter (darker) gray curve: the ratio of rewarded to sent
(rewarded to downloaded) documents. Horizontal lines: average
values of the two curves during two workdays. Light (darker) gray
regions: number of foragers is 10 (number of foragers increases
from 14 to 19). \textbf{(b):} Forager population and forager
values vs. time. The scores of the foragers are given with
different gray levels. Starting of white horizontal line means
forager multiplication. (There was a break in our Internet
connection in the middle of the second week.)} \label{f:fig2}
\end{figure}

In the presented experiments we applied the cost/reward ratio
described in Section \ref{s:methods}. Larger cost/reward ratio
increased the foraging areas and we noted a sudden increase in
extinction probability.

Time-lag between publishing news and finding those decreases
already after a few days (see Fig. \ref{f:fig2a}): the ratio of
maxima to minima of the curves increases; and also, fewer news
published on Friday were picked up on Saturday, a day late, during
the second weekend of the experiment than during the first.
Further gains in downloading speed are indicated by the relative
shift between lighter gray and darker gray peaks. The darker gray
peaks keep their maxima at around midnight GMT, lighter gray peaks
shift to earlier times by about 6 hours. The shift is due to
changes in the number of sent documents. The minima of this number
shifts to around 6:00 P.M. GMT, when it is around 3:00 A.M. in
Japan. (Identical dates can be found for a 48 hour period centered
around noon GMT.) Maxima of the relevant documents are at around
11:00 P.M. GMT (around 6:00 P.M. EST of the US). During the first
week, the horizontal lines (average values of the corresponding
curves during 2 workdays) are very close to each other. Both
averages increase for the second week. The ratio of sent to
reinforced documents increases more. At the beginning of the
darker gray region, the relative shift of the two curves is large,
at the end it is small, but becomes large again after that region,
when multiplication slows down.

The multiplication of the foragers is shown in Fig.~\ref{f:fig2b}.
Gray levels of this figure represent the value of the foragers,
the range goes from 60 (darker) to 200 (lighter). In the time
region studied, the values of the foragers have never fallen below
60. Upon bipartition new individuals are separated by horizontal
white lines in the figure.

\subsection{Compartmentalization} \label{s:compartment}

\begin{figure}[t!]
  \centering
   \subfigure[Early development]
      {
         \includegraphics[width=4cm]{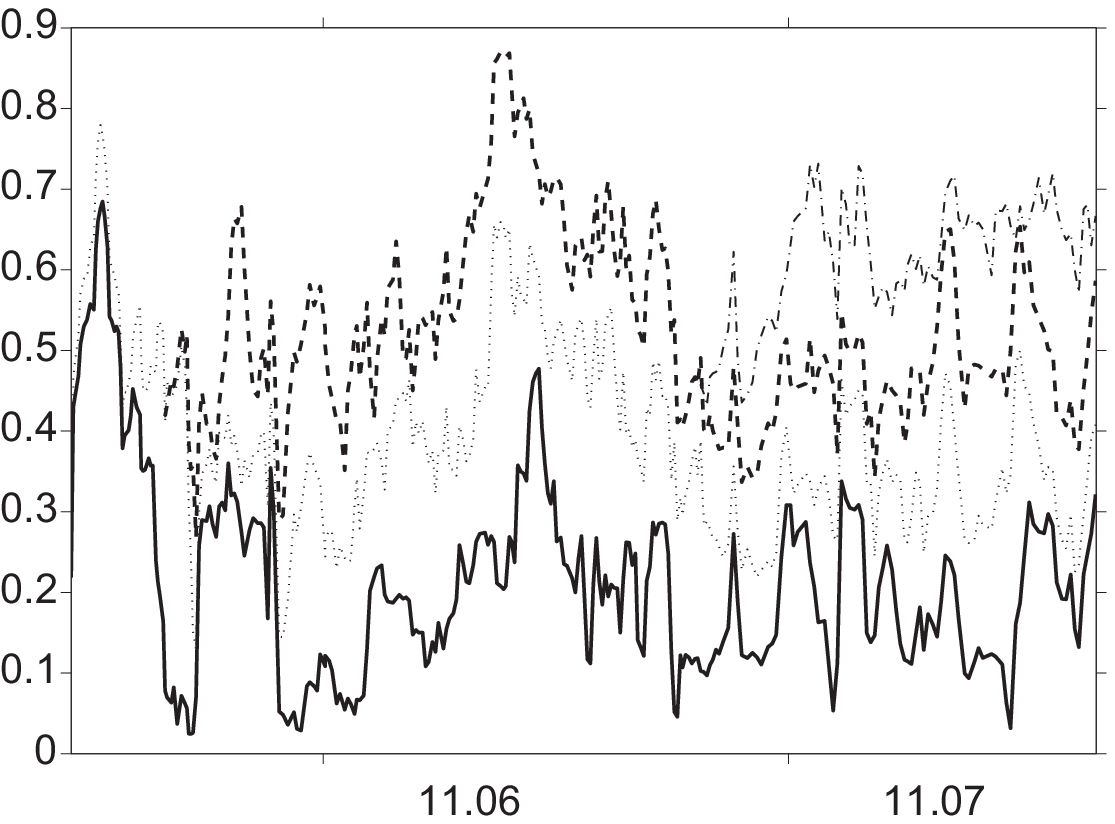}\label{f:fig3a}
      }
   \subfigure[Two week period]
      {
         \includegraphics[width=4cm]{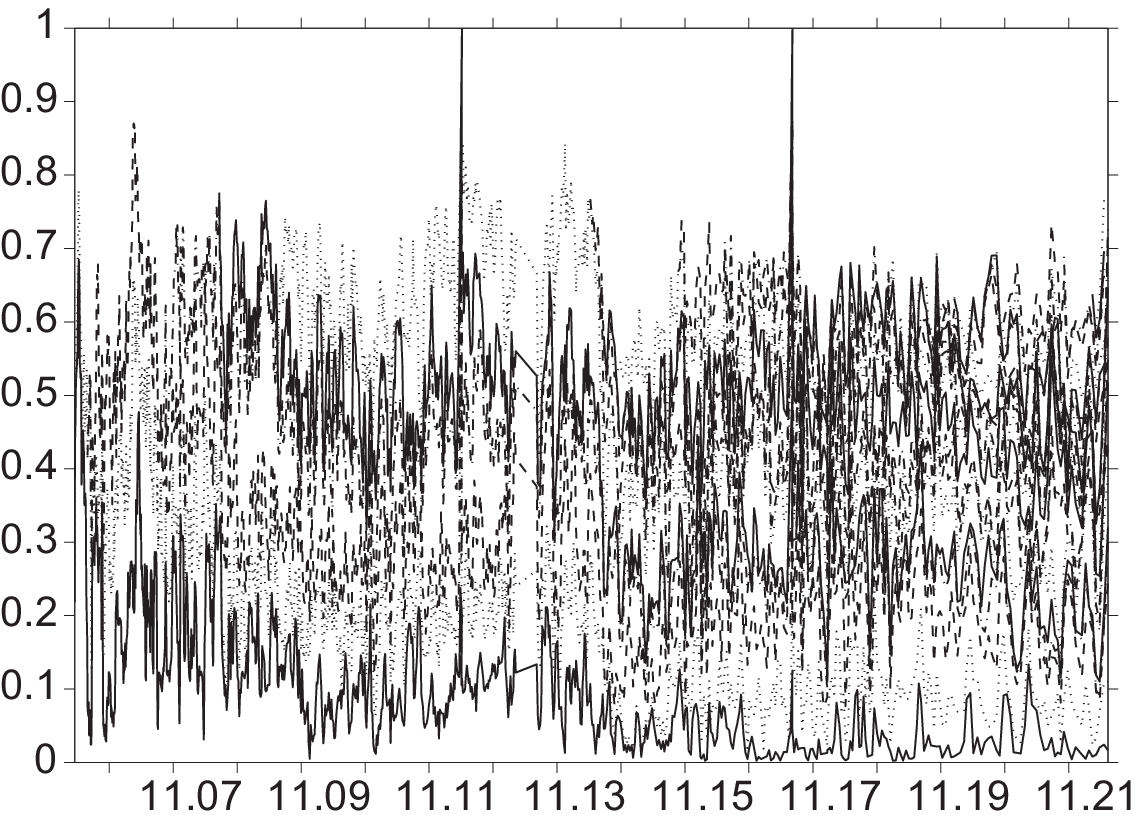}\label{f:fig3b}
      }
  \subfigure[Two step trajectories]
      {
         \includegraphics[width=4cm]{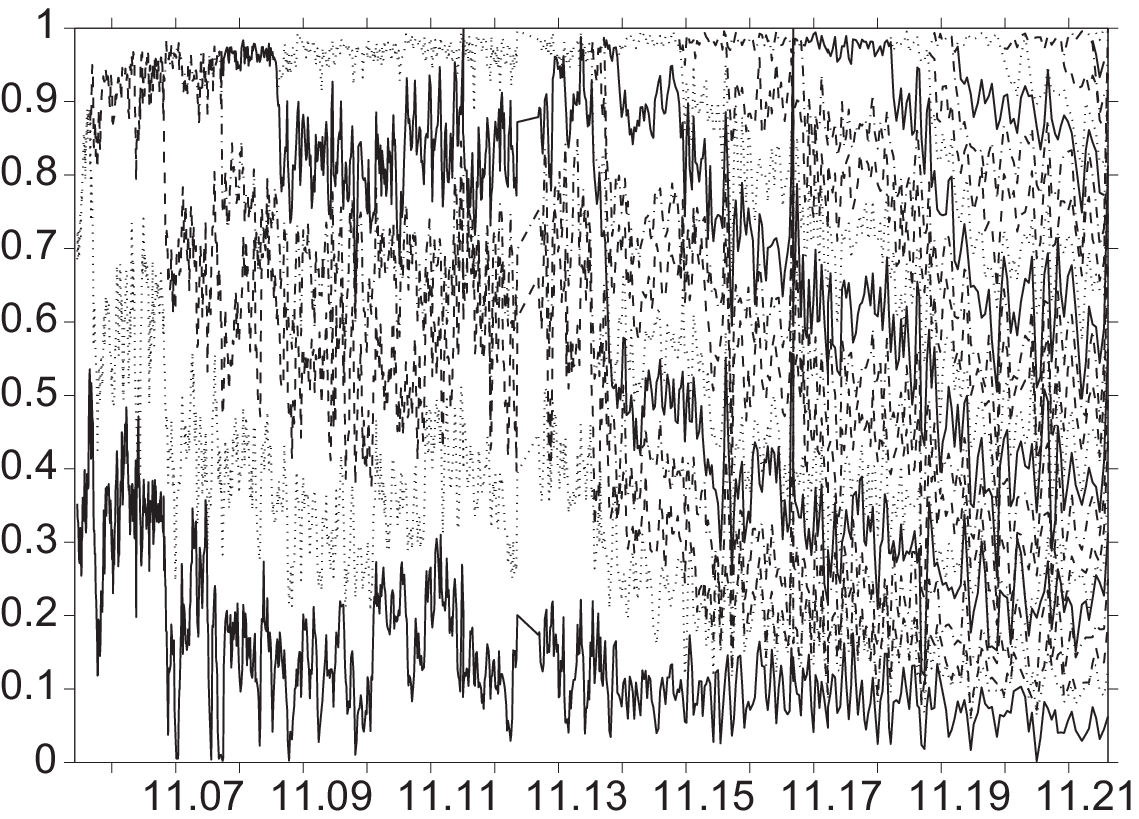}\label{f:fig3c}
      }
\caption{\textbf{Division of work} \newline Horizontal axis in
`month.day' units. \textbf{(a):} Number of sites visited by only
one forager relative to the number of all visited sites in a
finite time period ($\approx 75 mins$). Contribution of a single
forager is superimposed on cumulated contributions of older
foragers. The difference between 1.0 and the upper boundary of the
curves corresponds to the ratio of sites visited by more than one
forager. Duration: about three days. \textbf{(b):} Same for 16 day
period. \textbf{(c):} The ratio of different two step trajectories
relative to all two step trajectories conditioned that the two
step trajectories start from the same site, belong to different
foragers and are in a finite time period ($\approx 75 mins$).
Contribution of a single forager is superimposed on cumulated
contributions of older foragers. The difference between 1.0 and
the upper boundary of the curves corresponds to the ratio of the
conditioned 2 step trajectories taken by more than one forager.}
\label{f:fig3}
\end{figure}

Division of work is illustrated by Fig.~\ref{f:fig3}. According to
Fig.~\ref{f:fig3b} large proportion of the sites are visited
exclusively by not more than one forager. Only about 40\% of the
sites is visited by more than one forager. Figure \ref{f:fig3a}
demonstrates that new foragers occupy their territories quickly.
Figure \ref{f:fig3b} shows that similar data were found for few
(2-4) and for many (22) foragers (upper boundary is mainly between
0.6 and 0.7 throughout the figure). The figures depict the
contributions of individual foragers: as new foragers start they
quickly find good terrains while the older ones still keep their
good territories. The environment changes very quickly, cca. 1200
new URLs were found every day. Time intervals of about 75 mins
were investigated.

Figure ~\ref{f:fig3c} depicts the lack of overlap amongst forager
trajectories. A relatively large number of sites were visited by
different foragers. The question is if these foragers were
collecting similar or different news. If they had the same `goal'
then they presumably made similar decisions and, in turn, their
next step was the same. According to Figure ~\ref{f:fig3c} the
ratio of such 2 step trajectories -- i.e., the difference between
the upper cover of the curves and value 1.0 -- drops quickly at
the beginning of the experiment, it remains very small and, it
decreases further as the number of foragers is growing. Given that
the increase of the number of foragers gave rise to the decrease
of individual foraging time, the actual numbers are only
indicative. Nevertheless, the fast decrease at the beginning and
the small ratio for few as well as for many foragers provides
support that foragers followed different paths, that is, foragers
developed different behaviors. Differences between
hunting/foraging territories and/or differences between consumed
food are called compartmentalization (sometimes called niche
formation)
\cite{Clarck00dynamic,Csanyi89evolutionary,Fryxell98individual,kampis91selfmodifying}.
Figure \ref{f:fig3} demonstrates that compartmentalization, is
fast and efficient in our algorithm.

\section{Discussion and Summary}
\label{s:disc}

New Internet pages have been introduced continuously on the vast
Internet news domain that we studied. The population of news
foragers can be viewed as a rapidly self-assembling and adapting
news detector. The efficiency and speed of novelty detection is
increasing. This occurs in spite of the fact that the structure of
the environment is changing very quickly: the number of newly
discovered URLs was about constant versus time. Such drastic
changes are followed by the news detector, which continuously
reassembles itself and improves its monitoring efficiency.

In summary, our application demonstrates that evolutionary
algorithms can efficiently operate on scale-free networks. Similar
results can be expected in problems different from Internet
searches. However, we believe that the speed and the efficiency of
work sharing is mostly due to the highly clustered scale-free
small world structure of the Internet, a terrain full of traps.
These traps were turned into foraging-fields using the
two-component long-term memory, (a) the list of promising starting
points that we tentatively called `procedural memory' and (b) the
long-term profit (or reward) estimator. Both components are
needed: without (a) foragers can not escape traps, without (b)
forages can not compartmentalize, in other words share work with
high efficiency. The evolution of the forager population
progresses rapidly, making successful adaptation to fast-changing
worlds possible. These attractive properties are achieved without
direct communication among foragers, a major advantage in
communication networks.

We close by noting that the algorithm is not restricted to robotic
search on the Internet. Most notably, the algorithm seems
attractive for all networks where scale-free small world structure
is suspected, including social networks, cooperative networks as
well as others. (For a review, see, e.g.,
\cite{albert02statistical}.) Such human networks are constantly
developing over the Internet, see, e.g., the network of software
and hardware experts (http:\/\/www.experts-exchange.com\/).
Similar networks could be formed by human \textit{and} robotic
communities. To highlight the similarities, reward may mean money,
whereas multiplication could mean (i) hiring of employees, or,
(ii) purchasing computers and Internet bandwidth.

\subsubsection*{Acknowledgments}

This material is based upon work supported by the European Office
of Aerospace Research and Development, Air Force Office of
Scientific Research, Air Force Research Laboratory, under Contract
No. F61775-00-WE065. Any opinions, findings and conclusions or
recommendations expressed in this material are those of the
author(s) and do not necessarily reflect the views of the European
Office of Aerospace Research and Development, Air Force Office of
Scientific Research, Air Force Research Laboratory.

\small
\bibliographystyle{plain}
\bibliography{ecology}

\end{document}